\documentclass[%
reprint,
superscriptaddress,
amsmath,amssymb,
aps,
prb,
]{revtex4-2}

\usepackage{amsmath,amssymb,graphicx,bm}
\usepackage{xcolor}
\usepackage[colorlinks=true,urlcolor=blue,linkcolor=blue,citecolor=blue,bookmarks=false]{hyperref}
\usepackage{changes}
\def\la326{La$_3$Cd$_2$As$_6$}

\begin{document}
	
	\title{Putative excitonic insulating state in narrow-gap semiconductor La$_3$Cd$_2$As$_6$}

	\author{Caitlin S.~Kengle}
	\affiliation{Los Alamos National Laboratory, Los Alamos, New Mexico, 87545, USA}%
	
	\author{Noah Schnitzer}
	\affiliation{Department of Materials Science and Engineering, Cornell University, Ithaca, NY, 14853, USA}
	\affiliation{Kavli Institute at Cornell for Nanoscale Science, Ithaca, NY, 14853, USA}
	
	\author{Elizabeth A. Peterson}
	\affiliation{Los Alamos National Laboratory, Los Alamos, New Mexico, 87545, USA}%
	
	\author{Chunyu Guo}
	\affiliation{Max Planck Institute for the Structure and Dynamics of Matter, 22761 Hamburg, Germany}
	
	\author{Ling Zhang}
	\affiliation{Max Planck Institute for the Structure and Dynamics of Matter, 22761 Hamburg, Germany}
	
	\author{Matthew S. Cook}
	\affiliation{Los Alamos National Laboratory, Los Alamos, New Mexico, 87545, USA}
	
	\author{Jian-Xin Zhu}
	\affiliation{Los Alamos National Laboratory, Los Alamos, New Mexico, 87545, USA}%
	
	\author{Sean M. Thomas}
	\affiliation{Los Alamos National Laboratory, Los Alamos, New Mexico, 87545, USA}%

	\author{Philip J. W. Moll}
	\affiliation{Max Planck Institute for the Structure and Dynamics of Matter, 22761 Hamburg, Germany}
	
	\author{Filip Ronning}
	\affiliation{Los Alamos National Laboratory, Los Alamos, New Mexico, 87545, USA}%
	
	\author{Priscila F.S. Rosa}
	\affiliation{Los Alamos National Laboratory, Los Alamos, New Mexico, 87545, USA}%
	
	\date{\today}
	
	\begin{abstract}		
		Excitonic insulators are electronically-driven phases of matter characterized by the spontaneous condensation of electron-hole pairs. Here we show that \la326 undergoes a transition at $T_{0}=278$~K to a highly insulating state with no accompanying structural transition. We observe quasi-two-dimensional electrical transport and charge fluctuations consistent with an electronic transition enabled by enhanced Coulomb interactions. Density functional theory calculations are unable to replicate the insulating ground state. 
		Our results support the opening of a gap by excitonic effects at  $T_{0}$, placing \la326 as a rare example of a bulk excitonic insulator.
	\end{abstract}
	
	\maketitle

	The fascinating concept of an excitonic insulator (EI), wherein conduction-band electrons and valence-band holes form bound states known as excitons, was first proposed more than half a century ago \cite{Kohn1967PhysRev,Keldysh2024Book,Kohn1967PRL,Halperin1968Revmod}.
	When the binding energy of an exciton is greater than the semiconducting gap ($|E_B| > \Delta$), the material is unstable against the formation of excitons. 
	Depending on the carrier density, an EI state can be thought of as either  Bardeen-Cooper-Schrieffer-like, namely a condensate of loosely-bound electron-hole pairs, or Bose-Einstein condensate-like, when preformed excitons Bose-condense \cite{Kohn1967PRL, Chen2005PhysRep}.
	From an applied perspective, interest in exciton condensates stems from the potential for superfluid transport at moderate temperatures, which may enable minimally dissipative next-generation devices \cite{Li2017NatPhys, Gupta2020NatComm}. 
	
	The EI state has been theorized and experimentally realized in only a few two-dimensional systems, e.g., quantum Hall bilayer systems, monolayer transition metal dichalcogenides, and Moir\'e lattices \cite{Eisenstein2014AnnRev, Eisenstein2004Nature, Spielman2000PRL, Jia2022NatPhys}. In other cases, the state is argued to be present only under applied pressure \cite{Bucher1991PRL} or concomitant with a Peierls-type charge density wave (CDW) \cite{Cercellier2007PRL, Wakisaka2009PRL}.
	
	Although electronically driven, the EI transition is often accompanied by a structural transition with the same symmetry due to electron-phonon coupling \cite{Baldini2023PNAS, Zenker2014PRB, Kohn1967PhysRev}. As a result, unambiguous experimental identification of EI states remains a challenge.
	In bulk semimetals, candidate EI materials have been studied using advanced spectroscopy techniques which are capable of disentangling these lattice and electronic orders to rule in or out candidate materials.
	For example, momentum-resolved electron energy loss spectroscopy showed phonon softening in the absence of plasmon softening indicating the formation of an exciton condensate in semimetal $1T$-TiSe$_2$ \cite{Kogar2015Science}, whereas time-resolved angle-resolved photoemission spectroscopy (trARPES) found that the spontaneous symmetry breaking previously thought to be evidence of an EI transition in Ta$_2$NiSe$_5$ is a purely structural distortion \cite{Baldini2023PNAS}.

	The observation of an EI state in a bulk semiconductor remains rare.
	Conventional experimental techniques -- those that cannot distinguish electronic and lattice degrees of freedom independently -- can be used in semiconductors with weak electron-phonon coupling or wherein the electronic and structural transitions have a natural separation in energy scales.
	For example, ARPES, scanning tunneling microscopy, electron diffraction, and single-crystal x-ray diffraction (SC-XRD) were recently used to detect an electronic phase transition in Ta$_2$Pd$_3$Te$_5$ accompanied by a very weak structural transition that arises due to weak electron-phonon coupling \cite{Huang2024PRX,Zhang2024PRX}. Tight-binding calculations of the material including electron-hole Coulomb interactions confirm that electronic symmetry breaking opens a many-body gap.
	\la326, having a narrow gap of 105(1) meV \cite{Piva_2020_ACS}, presents as an ideal candidate to explore a new EI candidate.
	
	\begin{figure*}
		\centering
		\includegraphics[width=1\linewidth]{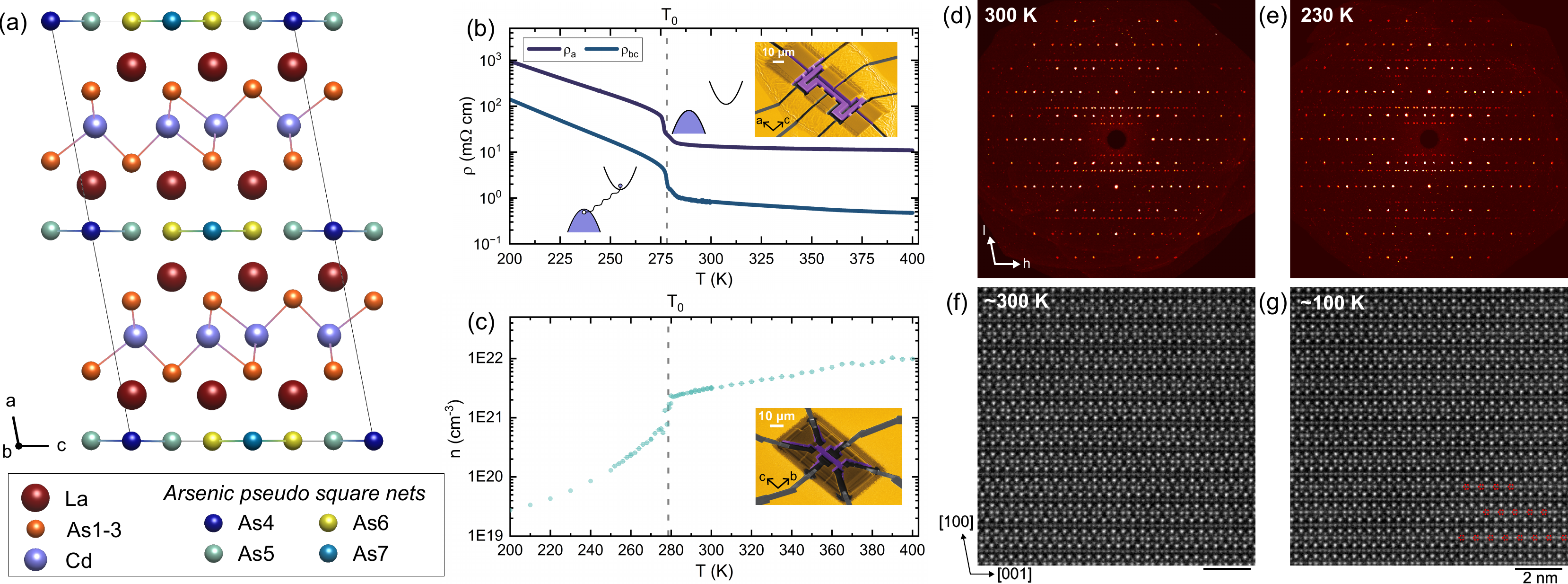}
		\caption{(a) Crystal structure of \la326 measured at 298~K. (b) Anisotropic electrical resistivity as a function of temperature in a microstructure fabricated using focused ion beam. Inset (photo): Microstructure used in anisotropic resistivity measurements. Inset (schematic): cartoon picture of the bands before and after an excitonic insulating transition. (c) Carrier density as a function of T from Hall measurements. Inset: Microstructure used in a Hall measurements. $h0\ell$ precession images taken at (d) 300 K and (e) 230 K. HAADF-STEM images acquired on the [010] zone axis at (f) 300 K and (g) 100 K with the corresponding log-transformed, Hann-filtered Fast Fourier Transform images at shown as insets. Dashed red circles in (g) shows a patch of partially occupied, nominally vacant Cd sites.}
		\label{fig:1}
	\end{figure*}

	Here we show evidence that narrow-gap semiconductor \la326 is an excitonic insulator below $T_0=278$K.
	The electrical resistivity is quasi- two-dimensional (2D), and in combination with Hall measurements, indicates a BEC-like insulator-to-excitonic insulator transition. 
	It does not undergo a structural transition between room temperature and $T = 100$ K, though structural refinements show evidence of electronic instability in the $bc$-plane via large anisotropic atomic displacements. 
	DFT calculations yield a robust metallic ground state regardless of perturbations to the structure or the inclusion of an on-site Coulomb repulsion term.
	Our combined results support that the electronic transition at $T_0$ is an EI transition driven by enhanced correlations due to the two-dimensional nature of the electronic structure. 
	Such electronic correlations, which are not considered in DFT, open the many-body gap at $T_0$. 
	
	Anisotropic resistivity ($\rho$) measurements were performed on a microstructure of \la326 shown in the inset of Fig.~\ref{fig:1}(b)]. 
	Both in-plane ($\rho_{bc}$) and out-of-plane ($\rho_{a}$) directions show a modest increase in $\rho$ on cooling from 400~K, a sharp jump at $T_0$, followed by a steeper increase below $T_0$, all of which indicate the opening of a gap at $T_0$ [see Fig.~\ref{fig:1}(b)]. 
	The in-plane resistivity, however, is about an order of magnitude smaller than the out-of-plane over the measured temperature range, indicating a quasi-2D electronic structure.

	The carrier density as a function of temperature derived from Hall measurements taken from 200 to 400 K is shown in Fig.~\ref{fig:1}(c) on the microstructure shown in the inset. 
	\la326 shows semiconducting behavior over the full temperature range with a significant change in the carrier concentration at $T_0$.
	Upon cooling, the carrier concentration decreases slightly, a factor-of-3 decrease at $T_0$, followed by a more rapid decrease below $T_0$.
    The consistency of this behavior with the quasi-2D transport are both evidence of opening of a gap with excitonic character.

	\begin{figure*}
		\centering
		\includegraphics[width=1\linewidth]{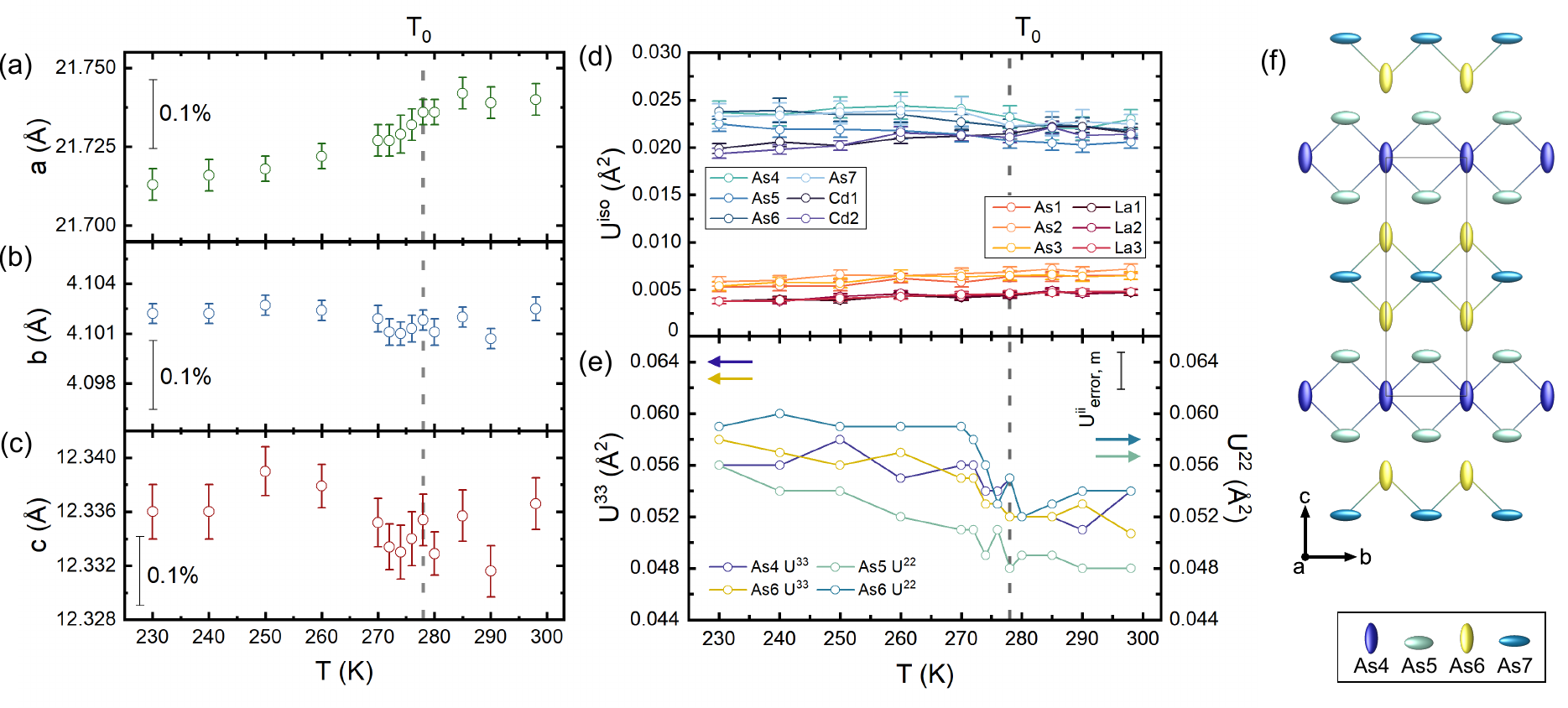}
		\caption{Temperature-dependent structural parameters. (a-c) Lattice parameters as a function of temperature. (d) Isotropic displacement parameters, U$^{\mathrm{iso}}$, as a function of temperature for all atoms. (e) Largest anisotropic ADPs for As4-7. (f) $bc$-plane showing the As4-7 atoms measured at 278 K. The atoms are represented with their 99\% probability ellipsoids. The pattern observed at 278 K is maintained throughout the temperature range.}
		\label{fig:lattice_adps}
	\end{figure*}
	
	SC-XRD is directly sensitive to the charge density of a material and can parse out information about the behavior of individual atoms in a crystal structure. The static atomic positions provide information about the localized equilibrium atomic positions, while the atomic displacement parameters (ADPs) provide information about the spatial extent, i.e., disorder and footprint of thermal vibration, of each atom in a crystal structure \cite{Trueblood1996IUC}. 
	Mapping the ADPs as a function of temperature in near an incipient phase transition can provide information about the dimensionality of the transition and of the material's electronic structure. 
	In contrast, techniques such as scanning tunneling electron microscopy (STEM) measurements can provide direct imaging of the atomic positions in a structure.
	
	At room temperature, our SC-XRD and STEM measurements confirm the monoclinic $C2/m$ structure of \la326 reported previously \cite{Piva_2020_ACS}. Figure \ref{fig:1}(a) shows the unit cell determined by SC-XRD with lattice parameters $ a = 21.740(5) \, \mathrm{\AA}$, $b = 4.1025(7) \, \mathrm{\AA}$, and $c = 12.3366(19) \, \mathrm{\AA}$.
	The precession images in the $h0\ell$ plane are presented in Fig.~\ref{fig:1}(d), wherein all peaks index to the unit cell. High-angle annular dark-field STEM (HAADF-STEM) imaging on the [010] zone axis, shown in Fig.~\ref{fig:1}(f,g), directly visualizes the atomic positions. 
	Our structural refinements find that the $y$-coordinates of all atoms in the structure are located at high symmetry positions such that static displacement along the crystallographic $b$-axis are prohibited.
	The HAADF-STEM images shown visualize the atoms in the $ac$-plane, ensuring that any potential static atomic shifts are captured. 
	
	Upon cooling below $T_0$, \la326 does not undergo any structural distortion that modifies the $C2/m$ symmetry. 
	Figure ~\ref{fig:1}(d,e) show the $h0\ell$ precession images taken above and below $T_0$ with no change in the Bragg peaks. 
	Slight differences intensities between the panels are attributable to the Debye-Waller effect.
	STEM images also taken above and below $T_0$ shown in Fig.~\ref{fig:1} (f,g) indicate no change in the real space crystal structure, additionally confirmed by SC-XRD structural refinements (Fig.~S1 \cite{supp}).

	Although no symmetry-breaking structural change occurs, the unit cell changes in this temperature range, outlined in Fig.~\ref{fig:lattice_adps}(a-c).  
	The $a$-axis remains constant within error above $T_0$, then shrinks suddenly upon cooling through $T_0$, reaching a reduction of $0.12\%$ at the lowest measured temperature, 230 K. 
	In the case of the $b$ and $c$ axes, any changes are masked by the experimental uncertainty.
	This leads to an overall volume contraction of about $0.14\%$ at 230 K, shown in Fig.~S2 \cite{supp}. 
	
	In addition to the slight changes in the lattice parameters, the ADPs reveal an instability in the As pseudo square nets which becomes enhanced at $T_0$.
	Figure \ref{fig:lattice_adps}(d) shows the isotropic ADPs for the 12 atomic sites in the structure as a function of temperature. 
	Temperature-dependent anisotropic ADPs for all sites can be found in Figs.~S3 \cite{supp}. 
	The red and orange empty circles correspond to the isotropic ADPs of the three La sites and As1-3 sites, respectively, which show a modest decrease with decreasing temperature, as would be expected from reduced thermal energy. 
	The purple and blue empty circles correspond to the isotropic ADPs of the Cd sites and the As4-7 sites, respectively, which exhibit values about four times greater than the La and As1-3 sites, which often indicate the presence of either static disorder or fluctuations.
	The large but decreasing-with-temperature Cd ADPs can be explained by site occupancy disorder. See Figs.~S4, S5, and S6 for additional details.

	Importantly, and in contrast to As4-5, there is no anomaly at $T_0$ for any of the Cd, La, or As1-3 sites.
	The arsenic pseudo-square nets exhibit a distortion pattern at room temperature which becomes enhanced at $T_0$, indicating an enhancement in an in-plane instability. Figure \ref{fig:lattice_adps}(f) shows a schematic of the $99\%$ displacement ellipsoids of the pseudo-square nets.
	Figure \ref{fig:lattice_adps}(e) shows the magnitude of the largest distortions of the As pseudo-square net: $U^{22}$ of As5 and As7 and $U^{33}$ of As4 and As6. 
	Above $T_0$, the values are approximately constant, but when the temperature is lowered through $T_0$, each of the parameters increases, with the largest ($U^{22}$ of As7) increasing by 15\% between $ T = 280$~K and $270$~K. 
	In contrast, the displacements along the $a$-axis for As4-7 are an order of magnitude smaller, indicating that the atomic motion is restricted to the $bc$ plane (see Fig.~S3 \cite{supp}). 
	The static atomic positions do not change significantly for any of the sites (Fig.~S1 \cite{supp}), revealing that the large ADPs arise as a result of fluctuations of the As4-7 atoms in the $bc$-plane.

\begin{figure}[t]
		\centering
		\includegraphics[width=1.0\linewidth]{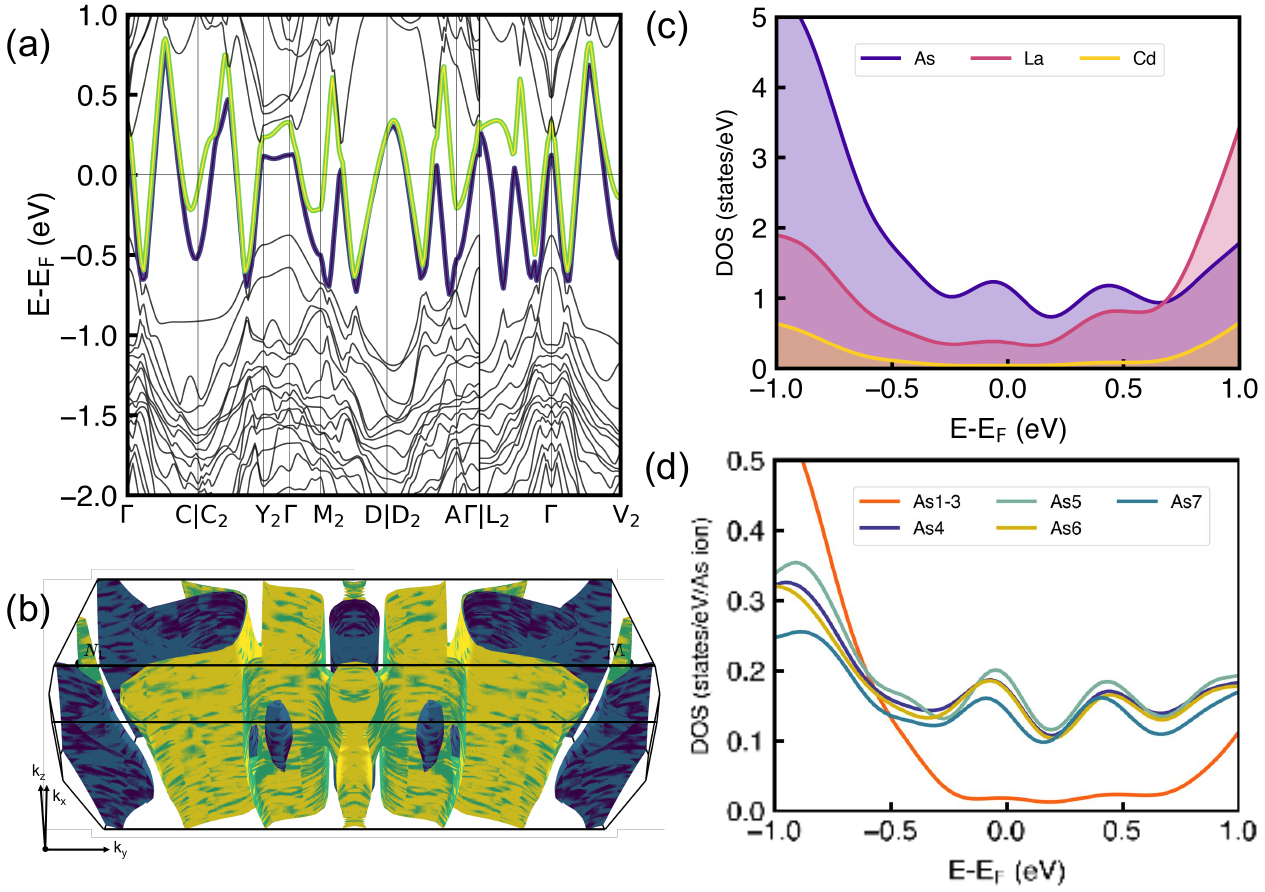}
		\caption{The electronic structure of the monoclinic $C2/m$ structure calculated using DFT with spin-orbit coupling. (a) The band structure, which exhibits highly dispersive bands near the Fermi energy. (b) Fermi surface calculated in the $C2/m$ parent cell showing the quasi-2D nature of the As4-7 $p$ bands. (c) The density of states (DOS) near the Fermi energy broken down by ion type dominated by As contributions. (d) The DOS near the Fermi energy per As site dominated by the As pseudo square net ions.}
		\label{fig:DFT}
	\end{figure}

	To understand the opening of a semiconducting gap in the absence of a structural transition, we turn to the band structure of \la326 calculated using DFT in the presence of spin-orbit coupling (SOC). As shown in Fig.~\ref{fig:DFT}(a), the band structure of the experimentally-determined $C2/m$ lattice is remarkably metallic and contains multiple highly dispersive bands crossing the Fermi energy. The resulting Fermi surface is shown in Fig.~\ref{fig:DFT}(b). Despite the sharp dispersions, the bands along the $k_x$ direction are nearly flat, in agreement with the quasi-2D electronic nature of \la326.

	Additional analysis was performed using DFT to further probe the relationship between the crystal structure and the electronic structure. The band structure for the parent $I4/mmm$ structure was also calculated, as shown in Fig.~S7 \cite{supp}, and reveals bands with qualitatively similar dispersion near the Fermi energy. The calculation was repeated in a relaxed parent $I4/mmm$ structure after introducing 33$\%$ Cd vacancies, but this did not qualitatively change the band dispersion near the Fermi energy, as shown in Fig.~S8(a) \cite{supp}. Additionally, the square net As ions in the $I4/mmm$ structure were rearranged to simulate a Peierls CDW-like distortion. Distortions were introduced in all three Cartesian directions and at varying degrees of deviation from the pristine square net geometry. These distortions also did not result in any substantial qualitative change of the band dispersion near the Fermi energy either, as shown in Fig.~S8(b) \cite{supp}. 
	
	Altogether, our calculations demonstrate that DFT alone is unable to open a band gap in \la326, suggesting the presence of electronic correlations. The contribution from each element to the DOS near the Fermi surface was calculated, shown in Fig.~\ref{fig:DFT}(c). At the Fermi level, the DOS is dominated by contributions from As. Importantly, the largest contribution near the Fermi energy comes from the pseudo square net As ions, As4-7, shown in the partial density of states in Figs.~\ref{fig:DFT}(d).
	As a result, one scenario to be considered is that of a Mott insulating transition due to correlation effects in the As pseudo square net layers.
	To determine whether DFT is erroneously predicting a metallic system because it fails to capture localization effects on the As pseudo square net ions, the DOS for the $C2/m$ structure was also calculated with variable Hubbard $U$ values (DFT+U), which includes on-site localization of the As $p$ orbitals. Our results reveal that the DOS near the Fermi energy only changes slightly with increasing As on-site localization (see Fig.~S9 \cite{supp}) and does not open a gap, indicating that the transition at $T_0$ seen in transport is not of a Mott insulator.
	
	Conventional symmetry-breaking phase transitions may arise from magnetic, lattice, or electronic instabilities.
	Because \la326 has no magnetic atoms, a magnetic transition can easily be excluded.
	In addition, our SC-XRD and STEM measurements reveal no symmetry-breaking structural phase transition at $T_0$, which leads us to the conclusion that the ground state of \la326 being insulating is electronic in origin

	Although clearly electronic, the exact type of electronic transition is not immediately apparent. 
	Our experimental results could in principle be interpreted as evidence of a Mott insulating state \cite{Mott1949PNAS, Phillips2010RevMod, Brandow1977AdvPhys}, as observed in transition metals, lanthanides, and actinides with partially filled bands \cite{Mott1949PNAS,Wen2013ACS,Nguyen2018PRM}. 
	This interpretation is ruled out by our DFT calculations where including on-site localization does not impact the electronic structure near the Fermi energy in any substantial way. 
	
	With the exclusion of a Mott insulating transition, our results are fully consistent with an excitonic insulating state.
	Hall measurements show a insulator to insulator transition at $T_0$.
	The electronic behavior of \la326 is 2D, with evidence of fluctuations in the As pseudo square nets in the $bc$-plane. 
	DFT predicts that the Fermi surface is strongly 2D and dominated by the As4-7 pseudo square net atoms.  
	Anisotropic resistivity measurements indicate that the conduction is nearly 10 times larger in plane compared to out-of-plane. 
	SC-XRD results show strong fluctuations of the electron density of the As pseudo square nets, indicating that there are strong correlations that get enhanced at $T_0$. 
	Taken together, our results are indicate that the ground state of \la326 is an electronically-driven excitonic insulating ground state, a first of its kind example of a BEC-insulator to excitonic insulator transition in a bulk material.

	\section{Acknowledgments}
	The authors acknowledge the significant contribution of the late Professor Lena F. Kourkoutis to this study.
	
	CSK thanks A. Schmidt for insightful discussions. Work at Los Alamos National Laboratory was performed under the auspices of the U.S. Department of Energy, Office of Basic Energy Sciences, Division of Materials Science and Engineering under project “Quantum Fluctuations in Narrow-Band Systems”. CSK and EAP acknowledge support from the Los Alamos Laboratory Directed Research and Development program. CSK gratefully acknowledges the support of the U.S. Department of Energy through the LANL/LDRD Program and the G. T. Seaborg Institute. 
	
	Scanning electron microscope were performed at the Center for Integrated Nanotechnologies (CINT), an Office of Science User Facility operated for the U.S. Department of Energy Office of Science. 
	Density functional theory calculations were also performed at CINT, in partnership with the LANL Institutional Computing Program. Further computations were performed at the National Energy Research Scientific Computing Center (NERSC), a U.S. Department of Energy Office of Science User Facility located at Lawrence Berkeley National Laboratory using NERSC awards ERCAP0020494 and ERCAP0028014.
	
	N. Schnitzer thanks B. H. Goodge and D. A. Muller for helpful discussions. This work made use of the electron microscopy facility of the Platform for the Accelerated Realization, Analysis, and Discovery of Interface Materials (PARADIM), which is supported by the National Science Foundation under Cooperative Agreement No. DMR-2039380. N.S. acknowledges support from the NSF GRFP under award number DGE-2139899. The authors acknowledge the use of facilities and instrumentation supported by NSF through the Cornell University Materials Research Science and Engineering Center DMR-1719875, a Helios FIB supported by NSF (DMR-1539918), and FEI Titan Themis 300 acquired through NSF-MRI-1429155, with additional support from Cornell University, the Weill Institute and the Kavli Institute at Cornell. 
	
	\section{Data availability}
	Data related to the electron microscopy characterization are available on the Platform for the Accelerated Realization, Analysis, and Discovery of Interface Materials (PARADIM) database at https://doi.org/10.34863/xxxxx.

	\section{Appendix I: Methods}
	\subsection*{Microstructure Measurements}
	The microstructures are fabricated using the focused-ion-beam (FIB) technique. Lamellae of \la326 were extracted from the crystal and transferred to a sapphire, followed by detailed fabrication with designed patterns. Low ohmic contacts for electric measurements were established through a high-power DC gold sputtering. Upon completion of the fabrication process, a thin carbon layer is deposited on the top of the device to increase its durability in air. 
	
	Electrical transport at temperatures above 250 K were carried out using an ac resistance bridge in a standard 4-point configuration. At lower temperatures, a programmable current source and nanovoltmeter were used in delta mode because of the large impedance of the sample. All electrical transport data was collected in a temperature-controlled cryostat equipped with a 9T magnet.

	\subsection*{X-Ray Diffraction}
	Single crystal X-ray diffraction (SC-XRD) experiments were performed using a Bruker D8 Venture on a sample of \la326 at 13 temperatures. Samples were grown through chemical vapor transport \cite{Piva_2020_ACS}. An Incoatec I$\mu$S microfocus source (Mo K-$\alpha$ radiation, $\lambda = 0.71073 \, \mathrm{\AA}$) was used as the radiation source. Data were collected using a PHOTON II CPAD area detector. 
	
	The sample temperature was controlled by a constant flow of cold N$_2$ gas via Oxford Cryosystems N-Helix.
	Data were collected between room temperature and 230 K. Each raw dataset was individually processed with Bruker SAINT software, including multi-scan absorption correction. The unit cell of each dataset was determined independently to quantify thermal contraction. The initial crystallographic model was obtained via the intrinsic phasing method in SHELXT. Least-squares refinements were performed using SHELXL2018 \cite{2015_Sheldrick_Acta}. 
	
	\subsection{Electron Microscopy}
	Scanning transmission electron microscopy (STEM) measurements were performed on cross-sectional lamellae prepared via focused ion beam (FIB) lift-out with a Thermo Fisher Helios G4 UX FIB. Cryogenic (100 K) and room temperature (300 K) high-angle annular dark-field (HAADF) STEM imaging were performed on a FEI/Thermo Fisher Titan Themis 300 CryoS/TEM with a Gatan 636 double tilt liquid nitrogen cooling holder operating at 300 kV with a 30 mrad probe convergence semi-angle and 15 pA probe current. For high-precision structural measurements, series of 40 rapid-frame images ($\sim$0.4 sec.~per frame) were acquired, aligned, and averaged by a method of rigid registration optimized to prevent lattice hops \cite{savitzky2018image} to recover high signal-to-noise ratio, high fidelity atomic resolution images. 4D-STEM nano-diffraction experiments were performed with a 0.5 mrad probe convergence semi-angle and 36 pA probe current, and an EMPAD-G2 detector \cite{philipp2022very}.

	\subsection{Density Functional Theory}
	First-principles density functional theory (DFT) calculations were performed using a plane-wave basis and projector augmented wave pseudopotentials ~\cite{Kresse1999} as implemented in the Vienna \textit{ab initio} simulation package (VASP) ~\cite{Kresse1996a,Kresse1996b}. The electronic structure was calculated in the generalized-gradient approximation (GGA) as implemented by Perdew, Burke, and Ernzerhof ~\cite{Perdew1996}. A 600 eV plane wave energy cutoff was used with a total energy convergence criteria of $10^{-5}$ eV. A $16\,\times16\,\times4$ $\Gamma$-centered $k$-point mesh was used for the $I4/mmm$ parent structure and a $6\,\times18\,\times4$ $\Gamma$-centered $k$-point mesh was used for the vacancy-ordered $C2/m$ structure. The experimental lattice parameters and ionic positions were used to calculate the electronic structure of both the $I4/mmm$ and $C2/m$ structures. Band structure and density of states calculations were performed including spin-orbit coupling (SOC) to account for relativistic effects. Further calculations were performed using an effective Hubbard U correction to the As $p$ orbitals to account for on-site localization effects, using $U= 1$, $2$, and $3$ eV, as implemented by Dudarev et al. ~\cite{Dudarev1998}. The Fermi surface for the C2/m structure in the conventional cell was plotted using IFermi \cite{Ganose2021}.
	
	\bibliography{bibliography}

\begin{thebibliography}{36}%
\makeatletter
\providecommand \@ifxundefined [1]{%
 \@ifx{#1\undefined}
}%
\providecommand \@ifnum [1]{%
 \ifnum #1\expandafter \@firstoftwo
 \else \expandafter \@secondoftwo
 \fi
}%
\providecommand \@ifx [1]{%
 \ifx #1\expandafter \@firstoftwo
 \else \expandafter \@secondoftwo
 \fi
}%
\providecommand \natexlab [1]{#1}%
\providecommand \enquote  [1]{``#1''}%
\providecommand \bibnamefont  [1]{#1}%
\providecommand \bibfnamefont [1]{#1}%
\providecommand \citenamefont [1]{#1}%
\providecommand \href@noop [0]{\@secondoftwo}%
\providecommand \href [0]{\begingroup \@sanitize@url \@href}%
\providecommand \@href[1]{\@@startlink{#1}\@@href}%
\providecommand \@@href[1]{\endgroup#1\@@endlink}%
\providecommand \@sanitize@url [0]{\catcode `\\12\catcode `\$12\catcode
  `\&12\catcode `\#12\catcode `\^12\catcode `\_12\catcode `\%12\relax}%
\providecommand \@@startlink[1]{}%
\providecommand \@@endlink[0]{}%
\providecommand \url  [0]{\begingroup\@sanitize@url \@url }%
\providecommand \@url [1]{\endgroup\@href {#1}{\urlprefix }}%
\providecommand \urlprefix  [0]{URL }%
\providecommand \Eprint [0]{\href }%
\providecommand \doibase [0]{https://doi.org/}%
\providecommand \selectlanguage [0]{\@gobble}%
\providecommand \bibinfo  [0]{\@secondoftwo}%
\providecommand \bibfield  [0]{\@secondoftwo}%
\providecommand \translation [1]{[#1]}%
\providecommand \BibitemOpen [0]{}%
\providecommand \bibitemStop [0]{}%
\providecommand \bibitemNoStop [0]{.\EOS\space}%
\providecommand \EOS [0]{\spacefactor3000\relax}%
\providecommand \BibitemShut  [1]{\csname bibitem#1\endcsname}%
\let\auto@bib@innerbib\@empty
\bibitem [{\citenamefont {J\'erome}\ \emph {et~al.}(1967)\citenamefont
  {J\'erome}, \citenamefont {Rice},\ and\ \citenamefont
  {Kohn}}]{Kohn1967PhysRev}%
  \BibitemOpen
  \bibfield  {author} {\bibinfo {author} {\bibfnamefont {D.}~\bibnamefont
  {J\'erome}}, \bibinfo {author} {\bibfnamefont {T.~M.}\ \bibnamefont {Rice}},\
  and\ \bibinfo {author} {\bibfnamefont {W.}~\bibnamefont {Kohn}},\ }\bibfield
  {title} {\bibinfo {title} {Excitonic insulator},\ }\href
  {https://doi.org/10.1103/PhysRev.158.462} {\bibfield  {journal} {\bibinfo
  {journal} {Phys. Rev.}\ }\textbf {\bibinfo {volume} {158}},\ \bibinfo {pages}
  {462} (\bibinfo {year} {1967})}\BibitemShut {NoStop}%
\bibitem [{\citenamefont {Keldysh}\ and\ \citenamefont
  {Kopaev}(2024)}]{Keldysh2024Book}%
  \BibitemOpen
  \bibfield  {author} {\bibinfo {author} {\bibfnamefont {L.}~\bibnamefont
  {Keldysh}}\ and\ \bibinfo {author} {\bibfnamefont {Y.~V.}\ \bibnamefont
  {Kopaev}},\ }\bibfield  {title} {\bibinfo {title} {Possible instability of
  the semimetallic state toward coulomb interaction},\ }in\ \href
  {https://doi.org/10.1142/9789811279461_0006} {\emph {\bibinfo {booktitle}
  {Selected Papers of Leonid V Keldysh}}}\ (\bibinfo  {publisher} {World
  Scientific},\ \bibinfo {year} {2024})\ pp.\ \bibinfo {pages}
  {41--46}\BibitemShut {NoStop}%
\bibitem [{\citenamefont {Kohn}(1967)}]{Kohn1967PRL}%
  \BibitemOpen
  \bibfield  {author} {\bibinfo {author} {\bibfnamefont {W.}~\bibnamefont
  {Kohn}},\ }\bibfield  {title} {\bibinfo {title} {Excitonic phases},\ }\href
  {https://doi.org/10.1103/PhysRevLett.19.439} {\bibfield  {journal} {\bibinfo
  {journal} {Phys. Rev. Lett.}\ }\textbf {\bibinfo {volume} {19}},\ \bibinfo
  {pages} {439} (\bibinfo {year} {1967})}\BibitemShut {NoStop}%
\bibitem [{\citenamefont {Halperin}\ and\ \citenamefont
  {Rice}(1968)}]{Halperin1968Revmod}%
  \BibitemOpen
  \bibfield  {author} {\bibinfo {author} {\bibfnamefont {B.~I.}\ \bibnamefont
  {Halperin}}\ and\ \bibinfo {author} {\bibfnamefont {T.~M.}\ \bibnamefont
  {Rice}},\ }\bibfield  {title} {\bibinfo {title} {Possible anomalies at a
  semimetal-semiconductor transistion},\ }\href
  {https://doi.org/10.1103/RevModPhys.40.755} {\bibfield  {journal} {\bibinfo
  {journal} {Rev. Mod. Phys.}\ }\textbf {\bibinfo {volume} {40}},\ \bibinfo
  {pages} {755} (\bibinfo {year} {1968})}\BibitemShut {NoStop}%
\bibitem [{\citenamefont {Chen}\ \emph {et~al.}(2005)\citenamefont {Chen},
  \citenamefont {Stajic}, \citenamefont {Tan},\ and\ \citenamefont
  {Levin}}]{Chen2005PhysRep}%
  \BibitemOpen
  \bibfield  {author} {\bibinfo {author} {\bibfnamefont {Q.}~\bibnamefont
  {Chen}}, \bibinfo {author} {\bibfnamefont {J.}~\bibnamefont {Stajic}},
  \bibinfo {author} {\bibfnamefont {S.}~\bibnamefont {Tan}},\ and\ \bibinfo
  {author} {\bibfnamefont {K.}~\bibnamefont {Levin}},\ }\bibfield  {title}
  {\bibinfo {title} {{BCS--BEC crossover: From high temperature superconductors
  to ultracold superfluids}},\ }\href
  {https://doi.org/10.1016/j.physrep.2005.02.005} {\bibfield  {journal}
  {\bibinfo  {journal} {Physics Reports}\ }\textbf {\bibinfo {volume} {412}},\
  \bibinfo {pages} {1} (\bibinfo {year} {2005})}\BibitemShut {NoStop}%
\bibitem [{\citenamefont {Li}\ \emph {et~al.}(2017)\citenamefont {Li},
  \citenamefont {Taniguchi}, \citenamefont {Watanabe}, \citenamefont {Hone},\
  and\ \citenamefont {Dean}}]{Li2017NatPhys}%
  \BibitemOpen
  \bibfield  {author} {\bibinfo {author} {\bibfnamefont {J.}~\bibnamefont
  {Li}}, \bibinfo {author} {\bibfnamefont {T.}~\bibnamefont {Taniguchi}},
  \bibinfo {author} {\bibfnamefont {K.}~\bibnamefont {Watanabe}}, \bibinfo
  {author} {\bibfnamefont {J.}~\bibnamefont {Hone}},\ and\ \bibinfo {author}
  {\bibfnamefont {C.}~\bibnamefont {Dean}},\ }\bibfield  {title} {\bibinfo
  {title} {Excitonic superfluid phase in double bilayer graphene},\ }\href
  {https://doi.org/10.1038/nphys4140} {\bibfield  {journal} {\bibinfo
  {journal} {Nature Physics}\ }\textbf {\bibinfo {volume} {13}},\ \bibinfo
  {pages} {751} (\bibinfo {year} {2017})}\BibitemShut {NoStop}%
\bibitem [{\citenamefont {Gupta}\ \emph {et~al.}(2020)\citenamefont {Gupta},
  \citenamefont {Kutana},\ and\ \citenamefont {Yakobson}}]{Gupta2020NatComm}%
  \BibitemOpen
  \bibfield  {author} {\bibinfo {author} {\bibfnamefont {S.}~\bibnamefont
  {Gupta}}, \bibinfo {author} {\bibfnamefont {A.}~\bibnamefont {Kutana}},\ and\
  \bibinfo {author} {\bibfnamefont {B.~I.}\ \bibnamefont {Yakobson}},\
  }\bibfield  {title} {\bibinfo {title} {Heterobilayers of 2d materials as a
  platform for excitonic superfluidity},\ }\href
  {https://doi.org/10.1038/s41467-020-16737-0} {\bibfield  {journal} {\bibinfo
  {journal} {Nature Communications}\ }\textbf {\bibinfo {volume} {11}},\
  \bibinfo {pages} {2989} (\bibinfo {year} {2020})}\BibitemShut {NoStop}%
\bibitem [{\citenamefont {Eisenstein}(2014)}]{Eisenstein2014AnnRev}%
  \BibitemOpen
  \bibfield  {author} {\bibinfo {author} {\bibfnamefont {J.}~\bibnamefont
  {Eisenstein}},\ }\bibfield  {title} {\bibinfo {title} {Exciton condensation
  in bilayer quantum hall systems},\ }\href
  {https://doi.org/0.1146/annurev-conmatphys-031113-133832} {\bibfield
  {journal} {\bibinfo  {journal} {Annu. Rev. Condens. Matter Phys.}\ }\textbf
  {\bibinfo {volume} {5}},\ \bibinfo {pages} {159} (\bibinfo {year}
  {2014})}\BibitemShut {NoStop}%
\bibitem [{\citenamefont {Eisenstein}\ and\ \citenamefont
  {MacDonald}(2004)}]{Eisenstein2004Nature}%
  \BibitemOpen
  \bibfield  {author} {\bibinfo {author} {\bibfnamefont {J.}~\bibnamefont
  {Eisenstein}}\ and\ \bibinfo {author} {\bibfnamefont {A.~H.}\ \bibnamefont
  {MacDonald}},\ }\bibfield  {title} {\bibinfo {title} {Bose--einstein
  condensation of excitons in bilayer electron systems},\ }\href
  {https://doi.org/10.1038/nature03081} {\bibfield  {journal} {\bibinfo
  {journal} {Nature}\ }\textbf {\bibinfo {volume} {432}},\ \bibinfo {pages}
  {691} (\bibinfo {year} {2004})}\BibitemShut {NoStop}%
\bibitem [{\citenamefont {Spielman}\ \emph {et~al.}(2000)\citenamefont
  {Spielman}, \citenamefont {Eisenstein}, \citenamefont {Pfeiffer},\ and\
  \citenamefont {West}}]{Spielman2000PRL}%
  \BibitemOpen
  \bibfield  {author} {\bibinfo {author} {\bibfnamefont {I.~B.}\ \bibnamefont
  {Spielman}}, \bibinfo {author} {\bibfnamefont {J.~P.}\ \bibnamefont
  {Eisenstein}}, \bibinfo {author} {\bibfnamefont {L.~N.}\ \bibnamefont
  {Pfeiffer}},\ and\ \bibinfo {author} {\bibfnamefont {K.~W.}\ \bibnamefont
  {West}},\ }\bibfield  {title} {\bibinfo {title} {Resonantly enhanced
  tunneling in a double layer quantum hall ferromagnet},\ }\href
  {https://doi.org/10.1103/PhysRevLett.84.5808} {\bibfield  {journal} {\bibinfo
   {journal} {Phys. Rev. Lett.}\ }\textbf {\bibinfo {volume} {84}},\ \bibinfo
  {pages} {5808} (\bibinfo {year} {2000})}\BibitemShut {NoStop}%
\bibitem [{\citenamefont {Jia}\ \emph {et~al.}(2022)\citenamefont {Jia},
  \citenamefont {Wang}, \citenamefont {Chiu}, \citenamefont {Song},
  \citenamefont {Yu}, \citenamefont {J{\"a}ck}, \citenamefont {Lei},
  \citenamefont {Klemenz}, \citenamefont {Cevallos}, \citenamefont {Onyszczak},
  \citenamefont {Fishchenko}, \citenamefont {Liu}, \citenamefont {Farahi},
  \citenamefont {Xie}, \citenamefont {Xu}, \citenamefont {Watanabe},
  \citenamefont {Taniguichi}, \citenamefont {Bernevig}, \citenamefont {Cava},
  \citenamefont {Schoop}, \citenamefont {Yazdani},\ and\ \citenamefont
  {Wu}}]{Jia2022NatPhys}%
  \BibitemOpen
  \bibfield  {author} {\bibinfo {author} {\bibfnamefont {Y.}~\bibnamefont
  {Jia}}, \bibinfo {author} {\bibfnamefont {P.}~\bibnamefont {Wang}}, \bibinfo
  {author} {\bibfnamefont {C.-L.}\ \bibnamefont {Chiu}}, \bibinfo {author}
  {\bibfnamefont {Z.}~\bibnamefont {Song}}, \bibinfo {author} {\bibfnamefont
  {G.}~\bibnamefont {Yu}}, \bibinfo {author} {\bibfnamefont {B.}~\bibnamefont
  {J{\"a}ck}}, \bibinfo {author} {\bibfnamefont {S.}~\bibnamefont {Lei}},
  \bibinfo {author} {\bibfnamefont {S.}~\bibnamefont {Klemenz}}, \bibinfo
  {author} {\bibfnamefont {F.~A.}\ \bibnamefont {Cevallos}}, \bibinfo {author}
  {\bibfnamefont {M.}~\bibnamefont {Onyszczak}}, \bibinfo {author}
  {\bibfnamefont {N.}~\bibnamefont {Fishchenko}}, \bibinfo {author}
  {\bibfnamefont {X.}~\bibnamefont {Liu}}, \bibinfo {author} {\bibfnamefont
  {G.}~\bibnamefont {Farahi}}, \bibinfo {author} {\bibfnamefont
  {F.}~\bibnamefont {Xie}}, \bibinfo {author} {\bibfnamefont {Y.}~\bibnamefont
  {Xu}}, \bibinfo {author} {\bibfnamefont {K.}~\bibnamefont {Watanabe}},
  \bibinfo {author} {\bibfnamefont {T.}~\bibnamefont {Taniguichi}}, \bibinfo
  {author} {\bibfnamefont {B.~A.}\ \bibnamefont {Bernevig}}, \bibinfo {author}
  {\bibfnamefont {R.~J.}\ \bibnamefont {Cava}}, \bibinfo {author}
  {\bibfnamefont {L.~M.}\ \bibnamefont {Schoop}}, \bibinfo {author}
  {\bibfnamefont {A.}~\bibnamefont {Yazdani}},\ and\ \bibinfo {author}
  {\bibfnamefont {S.}~\bibnamefont {Wu}},\ }\bibfield  {title} {\bibinfo
  {title} {Evidence for a monolayer excitonic insulator},\ }\href
  {https://doi.org/10.1038/s41567-021-01422-w} {\bibfield  {journal} {\bibinfo
  {journal} {Nature Physics}\ }\textbf {\bibinfo {volume} {18}},\ \bibinfo
  {pages} {87} (\bibinfo {year} {2022})}\BibitemShut {NoStop}%
\bibitem [{\citenamefont {Bucher}\ \emph {et~al.}(1991)\citenamefont {Bucher},
  \citenamefont {Steiner},\ and\ \citenamefont {Wachter}}]{Bucher1991PRL}%
  \BibitemOpen
  \bibfield  {author} {\bibinfo {author} {\bibfnamefont {B.}~\bibnamefont
  {Bucher}}, \bibinfo {author} {\bibfnamefont {P.}~\bibnamefont {Steiner}},\
  and\ \bibinfo {author} {\bibfnamefont {P.}~\bibnamefont {Wachter}},\
  }\bibfield  {title} {\bibinfo {title} {{Excitonic insulator phase in
  TmSe$_{0.45}$Te$_{0.55}$}},\ }\href
  {https://doi.org/10.1103/PhysRevLett.67.2717} {\bibfield  {journal} {\bibinfo
   {journal} {Phys. Rev. Lett.}\ }\textbf {\bibinfo {volume} {67}},\ \bibinfo
  {pages} {2717} (\bibinfo {year} {1991})}\BibitemShut {NoStop}%
\bibitem [{\citenamefont {Cercellier}\ \emph {et~al.}(2007)\citenamefont
  {Cercellier}, \citenamefont {Monney}, \citenamefont {Clerc}, \citenamefont
  {Battaglia}, \citenamefont {Despont}, \citenamefont {Garnier}, \citenamefont
  {Beck}, \citenamefont {Aebi}, \citenamefont {Patthey}, \citenamefont
  {Berger},\ and\ \citenamefont {Forr\'o}}]{Cercellier2007PRL}%
  \BibitemOpen
  \bibfield  {author} {\bibinfo {author} {\bibfnamefont {H.}~\bibnamefont
  {Cercellier}}, \bibinfo {author} {\bibfnamefont {C.}~\bibnamefont {Monney}},
  \bibinfo {author} {\bibfnamefont {F.}~\bibnamefont {Clerc}}, \bibinfo
  {author} {\bibfnamefont {C.}~\bibnamefont {Battaglia}}, \bibinfo {author}
  {\bibfnamefont {L.}~\bibnamefont {Despont}}, \bibinfo {author} {\bibfnamefont
  {M.~G.}\ \bibnamefont {Garnier}}, \bibinfo {author} {\bibfnamefont
  {H.}~\bibnamefont {Beck}}, \bibinfo {author} {\bibfnamefont {P.}~\bibnamefont
  {Aebi}}, \bibinfo {author} {\bibfnamefont {L.}~\bibnamefont {Patthey}},
  \bibinfo {author} {\bibfnamefont {H.}~\bibnamefont {Berger}},\ and\ \bibinfo
  {author} {\bibfnamefont {L.}~\bibnamefont {Forr\'o}},\ }\bibfield  {title}
  {\bibinfo {title} {{Evidence for an Excitonic Insulator Phase in
  1T-TiSe$_{2}$}},\ }\href {https://doi.org/10.1103/PhysRevLett.99.146403}
  {\bibfield  {journal} {\bibinfo  {journal} {Phys. Rev. Lett.}\ }\textbf
  {\bibinfo {volume} {99}},\ \bibinfo {pages} {146403} (\bibinfo {year}
  {2007})}\BibitemShut {NoStop}%
\bibitem [{\citenamefont {Wakisaka}\ \emph {et~al.}(2009)\citenamefont
  {Wakisaka}, \citenamefont {Sudayama}, \citenamefont {Takubo}, \citenamefont
  {Mizokawa}, \citenamefont {Arita}, \citenamefont {Namatame}, \citenamefont
  {Taniguchi}, \citenamefont {Katayama}, \citenamefont {Nohara},\ and\
  \citenamefont {Takagi}}]{Wakisaka2009PRL}%
  \BibitemOpen
  \bibfield  {author} {\bibinfo {author} {\bibfnamefont {Y.}~\bibnamefont
  {Wakisaka}}, \bibinfo {author} {\bibfnamefont {T.}~\bibnamefont {Sudayama}},
  \bibinfo {author} {\bibfnamefont {K.}~\bibnamefont {Takubo}}, \bibinfo
  {author} {\bibfnamefont {T.}~\bibnamefont {Mizokawa}}, \bibinfo {author}
  {\bibfnamefont {M.}~\bibnamefont {Arita}}, \bibinfo {author} {\bibfnamefont
  {H.}~\bibnamefont {Namatame}}, \bibinfo {author} {\bibfnamefont
  {M.}~\bibnamefont {Taniguchi}}, \bibinfo {author} {\bibfnamefont
  {N.}~\bibnamefont {Katayama}}, \bibinfo {author} {\bibfnamefont
  {M.}~\bibnamefont {Nohara}},\ and\ \bibinfo {author} {\bibfnamefont
  {H.}~\bibnamefont {Takagi}},\ }\bibfield  {title} {\bibinfo {title}
  {{Excitonic Insulator State in Ta$_{2}$NiSe$_{5}$ Probed by Photoemission
  Spectroscopy}},\ }\href {https://doi.org/10.1103/PhysRevLett.103.026402}
  {\bibfield  {journal} {\bibinfo  {journal} {Phys. Rev. Lett.}\ }\textbf
  {\bibinfo {volume} {103}},\ \bibinfo {pages} {026402} (\bibinfo {year}
  {2009})}\BibitemShut {NoStop}%
\bibitem [{\citenamefont {Baldini}\ \emph {et~al.}(2023)\citenamefont
  {Baldini}, \citenamefont {Zong}, \citenamefont {Choi}, \citenamefont {Lee},
  \citenamefont {Michael}, \citenamefont {Windgaetter}, \citenamefont {Mazin},
  \citenamefont {Latini}, \citenamefont {Azoury}, \citenamefont {Lv},
  \citenamefont {Kogar}, \citenamefont {Su}, \citenamefont {Wang},
  \citenamefont {Lu}, \citenamefont {Takayama}, \citenamefont {Takagi},
  \citenamefont {Millis}, \citenamefont {Rubio}, \citenamefont {Demler},\ and\
  \citenamefont {Gedik}}]{Baldini2023PNAS}%
  \BibitemOpen
  \bibfield  {author} {\bibinfo {author} {\bibfnamefont {E.}~\bibnamefont
  {Baldini}}, \bibinfo {author} {\bibfnamefont {A.}~\bibnamefont {Zong}},
  \bibinfo {author} {\bibfnamefont {D.}~\bibnamefont {Choi}}, \bibinfo {author}
  {\bibfnamefont {C.}~\bibnamefont {Lee}}, \bibinfo {author} {\bibfnamefont
  {M.~H.}\ \bibnamefont {Michael}}, \bibinfo {author} {\bibfnamefont
  {L.}~\bibnamefont {Windgaetter}}, \bibinfo {author} {\bibfnamefont {I.~I.}\
  \bibnamefont {Mazin}}, \bibinfo {author} {\bibfnamefont {S.}~\bibnamefont
  {Latini}}, \bibinfo {author} {\bibfnamefont {D.}~\bibnamefont {Azoury}},
  \bibinfo {author} {\bibfnamefont {B.}~\bibnamefont {Lv}}, \bibinfo {author}
  {\bibfnamefont {A.}~\bibnamefont {Kogar}}, \bibinfo {author} {\bibfnamefont
  {Y.}~\bibnamefont {Su}}, \bibinfo {author} {\bibfnamefont {Y.}~\bibnamefont
  {Wang}}, \bibinfo {author} {\bibfnamefont {Y.}~\bibnamefont {Lu}}, \bibinfo
  {author} {\bibfnamefont {T.}~\bibnamefont {Takayama}}, \bibinfo {author}
  {\bibfnamefont {H.}~\bibnamefont {Takagi}}, \bibinfo {author} {\bibfnamefont
  {A.~J.}\ \bibnamefont {Millis}}, \bibinfo {author} {\bibfnamefont
  {A.}~\bibnamefont {Rubio}}, \bibinfo {author} {\bibfnamefont
  {E.}~\bibnamefont {Demler}},\ and\ \bibinfo {author} {\bibfnamefont
  {N.}~\bibnamefont {Gedik}},\ }\bibfield  {title} {\bibinfo {title} {{The
  spontaneous symmetry breaking in Ta$_2$NiSe$_5$ is structural in nature}},\
  }\href {https://doi.org/10.1073/pnas.2221688120} {\bibfield  {journal}
  {\bibinfo  {journal} {Proceedings of the National Academy of Sciences}\
  }\textbf {\bibinfo {volume} {120}},\ \bibinfo {pages} {e2221688120} (\bibinfo
  {year} {2023})}\BibitemShut {NoStop}%
\bibitem [{\citenamefont {Zenker}\ \emph {et~al.}(2014)\citenamefont {Zenker},
  \citenamefont {Fehske},\ and\ \citenamefont {Beck}}]{Zenker2014PRB}%
  \BibitemOpen
  \bibfield  {author} {\bibinfo {author} {\bibfnamefont {B.}~\bibnamefont
  {Zenker}}, \bibinfo {author} {\bibfnamefont {H.}~\bibnamefont {Fehske}},\
  and\ \bibinfo {author} {\bibfnamefont {H.}~\bibnamefont {Beck}},\ }\bibfield
  {title} {\bibinfo {title} {Fate of the excitonic insulator in the presence of
  phonons},\ }\href {https://doi.org/10.1103/PhysRevB.90.195118} {\bibfield
  {journal} {\bibinfo  {journal} {Physical Review B}\ }\textbf {\bibinfo
  {volume} {90}},\ \bibinfo {pages} {195118} (\bibinfo {year}
  {2014})}\BibitemShut {NoStop}%
\bibitem [{\citenamefont {Kogar}\ \emph {et~al.}(2017)\citenamefont {Kogar},
  \citenamefont {Rak}, \citenamefont {Vig}, \citenamefont {Husain},
  \citenamefont {Flicker}, \citenamefont {Joe}, \citenamefont {Venema},
  \citenamefont {MacDougall}, \citenamefont {Chiang}, \citenamefont {Fradkin},
  \citenamefont {van Wezel},\ and\ \citenamefont
  {Abbamonte}}]{Kogar2015Science}%
  \BibitemOpen
  \bibfield  {author} {\bibinfo {author} {\bibfnamefont {A.}~\bibnamefont
  {Kogar}}, \bibinfo {author} {\bibfnamefont {M.~S.}\ \bibnamefont {Rak}},
  \bibinfo {author} {\bibfnamefont {S.}~\bibnamefont {Vig}}, \bibinfo {author}
  {\bibfnamefont {A.~A.}\ \bibnamefont {Husain}}, \bibinfo {author}
  {\bibfnamefont {F.}~\bibnamefont {Flicker}}, \bibinfo {author} {\bibfnamefont
  {Y.~I.}\ \bibnamefont {Joe}}, \bibinfo {author} {\bibfnamefont
  {L.}~\bibnamefont {Venema}}, \bibinfo {author} {\bibfnamefont {G.~J.}\
  \bibnamefont {MacDougall}}, \bibinfo {author} {\bibfnamefont {T.~C.}\
  \bibnamefont {Chiang}}, \bibinfo {author} {\bibfnamefont {E.}~\bibnamefont
  {Fradkin}}, \bibinfo {author} {\bibfnamefont {J.}~\bibnamefont {van Wezel}},\
  and\ \bibinfo {author} {\bibfnamefont {P.}~\bibnamefont {Abbamonte}},\
  }\bibfield  {title} {\bibinfo {title} {Signatures of exciton condensation in
  a transition metal dichalcogenide},\ }\href
  {https://doi.org/10.1126/science.aam6432} {\bibfield  {journal} {\bibinfo
  {journal} {Science}\ }\textbf {\bibinfo {volume} {358}},\ \bibinfo {pages}
  {1314} (\bibinfo {year} {2017})}\BibitemShut {NoStop}%
\bibitem [{\citenamefont {Huang}\ \emph {et~al.}(2024)\citenamefont {Huang},
  \citenamefont {Jiang}, \citenamefont {Yao}, \citenamefont {Yan},
  \citenamefont {Lei}, \citenamefont {Gao}, \citenamefont {Guo}, \citenamefont
  {Jin}, \citenamefont {Li}, \citenamefont {Yuan}, \citenamefont {Chai},
  \citenamefont {Sheng}, \citenamefont {Pan}, \citenamefont {Chen},
  \citenamefont {Liu}, \citenamefont {Gao}, \citenamefont {Qu}, \citenamefont
  {Liu}, \citenamefont {Jiang}, \citenamefont {Liu}, \citenamefont {Ma},
  \citenamefont {Zhou}, \citenamefont {Huang}, \citenamefont {Yun},
  \citenamefont {Zhang}, \citenamefont {Li}, \citenamefont {Jin}, \citenamefont
  {Ding}, \citenamefont {Shen}, \citenamefont {Su}, \citenamefont {Shi},
  \citenamefont {Wang},\ and\ \citenamefont {Qian}}]{Huang2024PRX}%
  \BibitemOpen
  \bibfield  {author} {\bibinfo {author} {\bibfnamefont {J.}~\bibnamefont
  {Huang}}, \bibinfo {author} {\bibfnamefont {B.}~\bibnamefont {Jiang}},
  \bibinfo {author} {\bibfnamefont {J.}~\bibnamefont {Yao}}, \bibinfo {author}
  {\bibfnamefont {D.}~\bibnamefont {Yan}}, \bibinfo {author} {\bibfnamefont
  {X.}~\bibnamefont {Lei}}, \bibinfo {author} {\bibfnamefont {J.}~\bibnamefont
  {Gao}}, \bibinfo {author} {\bibfnamefont {Z.}~\bibnamefont {Guo}}, \bibinfo
  {author} {\bibfnamefont {F.}~\bibnamefont {Jin}}, \bibinfo {author}
  {\bibfnamefont {Y.}~\bibnamefont {Li}}, \bibinfo {author} {\bibfnamefont
  {Z.}~\bibnamefont {Yuan}}, \bibinfo {author} {\bibfnamefont {C.}~\bibnamefont
  {Chai}}, \bibinfo {author} {\bibfnamefont {H.}~\bibnamefont {Sheng}},
  \bibinfo {author} {\bibfnamefont {M.}~\bibnamefont {Pan}}, \bibinfo {author}
  {\bibfnamefont {F.}~\bibnamefont {Chen}}, \bibinfo {author} {\bibfnamefont
  {J.}~\bibnamefont {Liu}}, \bibinfo {author} {\bibfnamefont {S.}~\bibnamefont
  {Gao}}, \bibinfo {author} {\bibfnamefont {G.}~\bibnamefont {Qu}}, \bibinfo
  {author} {\bibfnamefont {B.}~\bibnamefont {Liu}}, \bibinfo {author}
  {\bibfnamefont {Z.}~\bibnamefont {Jiang}}, \bibinfo {author} {\bibfnamefont
  {Z.}~\bibnamefont {Liu}}, \bibinfo {author} {\bibfnamefont {X.}~\bibnamefont
  {Ma}}, \bibinfo {author} {\bibfnamefont {S.}~\bibnamefont {Zhou}}, \bibinfo
  {author} {\bibfnamefont {Y.}~\bibnamefont {Huang}}, \bibinfo {author}
  {\bibfnamefont {C.}~\bibnamefont {Yun}}, \bibinfo {author} {\bibfnamefont
  {Q.}~\bibnamefont {Zhang}}, \bibinfo {author} {\bibfnamefont
  {S.}~\bibnamefont {Li}}, \bibinfo {author} {\bibfnamefont {S.}~\bibnamefont
  {Jin}}, \bibinfo {author} {\bibfnamefont {H.}~\bibnamefont {Ding}}, \bibinfo
  {author} {\bibfnamefont {J.}~\bibnamefont {Shen}}, \bibinfo {author}
  {\bibfnamefont {D.}~\bibnamefont {Su}}, \bibinfo {author} {\bibfnamefont
  {Y.}~\bibnamefont {Shi}}, \bibinfo {author} {\bibfnamefont {Z.}~\bibnamefont
  {Wang}},\ and\ \bibinfo {author} {\bibfnamefont {T.}~\bibnamefont {Qian}},\
  }\bibfield  {title} {\bibinfo {title} {{Evidence for an Excitonic Insulator
  State in Ta$_2$Pd$_3$Te$_5$}},\ }\href
  {https://doi.org/10.1103/PhysRevX.14.011046} {\bibfield  {journal} {\bibinfo
  {journal} {Physical Review X}\ }\textbf {\bibinfo {volume} {14}},\ \bibinfo
  {pages} {011046} (\bibinfo {year} {2024})}\BibitemShut {NoStop}%
\bibitem [{\citenamefont {Zhang}\ \emph {et~al.}(2024)\citenamefont {Zhang},
  \citenamefont {Dong}, \citenamefont {Yan}, \citenamefont {Jiang},
  \citenamefont {Yang}, \citenamefont {Li}, \citenamefont {Guo}, \citenamefont
  {Huang}, \citenamefont {Haobo}, \citenamefont {Li}, \citenamefont {Li},
  \citenamefont {Kurokawa}, \citenamefont {Wang}, \citenamefont {Nie},
  \citenamefont {Hashimoto}, \citenamefont {Lu}, \citenamefont {Jiao},
  \citenamefont {Shen}, \citenamefont {Qian}, \citenamefont {Wang},
  \citenamefont {Shi},\ and\ \citenamefont {Kondo}}]{Zhang2024PRX}%
  \BibitemOpen
  \bibfield  {author} {\bibinfo {author} {\bibfnamefont {P.}~\bibnamefont
  {Zhang}}, \bibinfo {author} {\bibfnamefont {Y.}~\bibnamefont {Dong}},
  \bibinfo {author} {\bibfnamefont {D.}~\bibnamefont {Yan}}, \bibinfo {author}
  {\bibfnamefont {B.}~\bibnamefont {Jiang}}, \bibinfo {author} {\bibfnamefont
  {T.}~\bibnamefont {Yang}}, \bibinfo {author} {\bibfnamefont {J.}~\bibnamefont
  {Li}}, \bibinfo {author} {\bibfnamefont {Z.}~\bibnamefont {Guo}}, \bibinfo
  {author} {\bibfnamefont {Y.}~\bibnamefont {Huang}}, \bibinfo {author}
  {\bibnamefont {Haobo}}, \bibinfo {author} {\bibfnamefont {Q.}~\bibnamefont
  {Li}}, \bibinfo {author} {\bibfnamefont {Y.}~\bibnamefont {Li}}, \bibinfo
  {author} {\bibfnamefont {K.}~\bibnamefont {Kurokawa}}, \bibinfo {author}
  {\bibfnamefont {R.}~\bibnamefont {Wang}}, \bibinfo {author} {\bibfnamefont
  {Y.}~\bibnamefont {Nie}}, \bibinfo {author} {\bibfnamefont {M.}~\bibnamefont
  {Hashimoto}}, \bibinfo {author} {\bibfnamefont {D.}~\bibnamefont {Lu}},
  \bibinfo {author} {\bibfnamefont {W.-H.}\ \bibnamefont {Jiao}}, \bibinfo
  {author} {\bibfnamefont {J.}~\bibnamefont {Shen}}, \bibinfo {author}
  {\bibfnamefont {T.}~\bibnamefont {Qian}}, \bibinfo {author} {\bibfnamefont
  {Z.}~\bibnamefont {Wang}}, \bibinfo {author} {\bibfnamefont {Y.}~\bibnamefont
  {Shi}},\ and\ \bibinfo {author} {\bibfnamefont {T.}~\bibnamefont {Kondo}},\
  }\bibfield  {title} {\bibinfo {title} {{Spontaneous Gap Opening and Potential
  Excitonic States in an Ideal Dirac Semimetal Ta$_{2}$Pd$_{3}$Te$_{5}$}},\
  }\href {https://doi.org/10.1103/PhysRevX.14.011047} {\bibfield  {journal}
  {\bibinfo  {journal} {Phys. Rev. X}\ }\textbf {\bibinfo {volume} {14}},\
  \bibinfo {pages} {011047} (\bibinfo {year} {2024})}\BibitemShut {NoStop}%
\bibitem [{\citenamefont {Trueblood}\ \emph {et~al.}(1996)\citenamefont
  {Trueblood}, \citenamefont {B{\"u}rgi}, \citenamefont {Burzlaff},
  \citenamefont {Dunitz}, \citenamefont {Gramaccioli}, \citenamefont {Schulz},
  \citenamefont {Shmueli},\ and\ \citenamefont {Abrahams}}]{Trueblood1996IUC}%
  \BibitemOpen
  \bibfield  {author} {\bibinfo {author} {\bibfnamefont {K.}~\bibnamefont
  {Trueblood}}, \bibinfo {author} {\bibfnamefont {H.-B.}\ \bibnamefont
  {B{\"u}rgi}}, \bibinfo {author} {\bibfnamefont {H.}~\bibnamefont {Burzlaff}},
  \bibinfo {author} {\bibfnamefont {J.}~\bibnamefont {Dunitz}}, \bibinfo
  {author} {\bibfnamefont {C.}~\bibnamefont {Gramaccioli}}, \bibinfo {author}
  {\bibfnamefont {H.}~\bibnamefont {Schulz}}, \bibinfo {author} {\bibfnamefont
  {U.}~\bibnamefont {Shmueli}},\ and\ \bibinfo {author} {\bibfnamefont
  {S.}~\bibnamefont {Abrahams}},\ }\bibfield  {title} {\bibinfo {title} {Atomic
  dispacement parameter nomenclature. report of a subcommittee on atomic
  displacement parameter nomenclature},\ }\href@noop {} {\bibfield  {journal}
  {\bibinfo  {journal} {Foundations of Crystallography}\ }\textbf {\bibinfo
  {volume} {52}},\ \bibinfo {pages} {770} (\bibinfo {year} {1996})}\BibitemShut
  {NoStop}%
\bibitem [{\citenamefont {Piva}\ \emph {et~al.}(2021)\citenamefont {Piva},
  \citenamefont {Rahn}, \citenamefont {Thomas}, \citenamefont {Scott},
  \citenamefont {Pagliuso}, \citenamefont {Thompson}, \citenamefont {Schoop},
  \citenamefont {Ronning},\ and\ \citenamefont {Rosa}}]{Piva_2020_ACS}%
  \BibitemOpen
  \bibfield  {author} {\bibinfo {author} {\bibfnamefont {M.~M.}\ \bibnamefont
  {Piva}}, \bibinfo {author} {\bibfnamefont {M.~C.}\ \bibnamefont {Rahn}},
  \bibinfo {author} {\bibfnamefont {S.~M.}\ \bibnamefont {Thomas}}, \bibinfo
  {author} {\bibfnamefont {B.~L.}\ \bibnamefont {Scott}}, \bibinfo {author}
  {\bibfnamefont {P.~G.}\ \bibnamefont {Pagliuso}}, \bibinfo {author}
  {\bibfnamefont {J.~D.}\ \bibnamefont {Thompson}}, \bibinfo {author}
  {\bibfnamefont {L.~M.}\ \bibnamefont {Schoop}}, \bibinfo {author}
  {\bibfnamefont {F.}~\bibnamefont {Ronning}},\ and\ \bibinfo {author}
  {\bibfnamefont {P.~F.~S.}\ \bibnamefont {Rosa}},\ }\bibfield  {title}
  {\bibinfo {title} {{Robust Narrow-Gap Semiconducting Behavior in Square-Net
  La$_3$Cd$_2$As$_6$}},\ }\href {https://doi.org/10.1021/acs.chemmater.1c00797}
  {\bibfield  {journal} {\bibinfo  {journal} {Chemistry of Materials}\ }\textbf
  {\bibinfo {volume} {33}},\ \bibinfo {pages} {4122} (\bibinfo {year}
  {2021})}\BibitemShut {NoStop}%
\bibitem [{sup()}]{supp}%
  \BibitemOpen
  \href@noop {} {\bibinfo {title} {{See Supplemental Material at xx for
  additional details on SC-XRD, cryo STEM, and DFT calculations.}}}\BibitemShut
  {Stop}%
\bibitem [{\citenamefont {Mott}(1949)}]{Mott1949PNAS}%
  \BibitemOpen
  \bibfield  {author} {\bibinfo {author} {\bibfnamefont {N.~F.}\ \bibnamefont
  {Mott}},\ }\bibfield  {title} {\bibinfo {title} {The basis of the electron
  theory of metals, with special reference to the transition metals},\ }\href
  {https://doi.org/10.1088/0370-1298/62/7/303} {\bibfield  {journal} {\bibinfo
  {journal} {Proceedings of the Physical Society. Section A}\ }\textbf
  {\bibinfo {volume} {62}},\ \bibinfo {pages} {416} (\bibinfo {year}
  {1949})}\BibitemShut {NoStop}%
\bibitem [{\citenamefont {Phillips}(2010)}]{Phillips2010RevMod}%
  \BibitemOpen
  \bibfield  {author} {\bibinfo {author} {\bibfnamefont {P.}~\bibnamefont
  {Phillips}},\ }\bibfield  {title} {\bibinfo {title} {{Colloquium: Identifying
  the propagating charge modes in doped Mott insulators}},\ }\href
  {https://doi.org/10.1103/RevModPhys.82.1719} {\bibfield  {journal} {\bibinfo
  {journal} {Reviews of Modern Physics}\ }\textbf {\bibinfo {volume} {82}},\
  \bibinfo {pages} {1719} (\bibinfo {year} {2010})}\BibitemShut {NoStop}%
\bibitem [{\citenamefont {Brandow}(1977)}]{Brandow1977AdvPhys}%
  \BibitemOpen
  \bibfield  {author} {\bibinfo {author} {\bibfnamefont {B.}~\bibnamefont
  {Brandow}},\ }\bibfield  {title} {\bibinfo {title} {{Electronic structure of
  Mott insulators}},\ }\href {https://doi.org/10.1080/00018737700101443}
  {\bibfield  {journal} {\bibinfo  {journal} {Advances in Physics}\ }\textbf
  {\bibinfo {volume} {26}},\ \bibinfo {pages} {651} (\bibinfo {year}
  {1977})}\BibitemShut {NoStop}%
\bibitem [{\citenamefont {Wen}\ \emph {et~al.}(2013)\citenamefont {Wen},
  \citenamefont {Martin}, \citenamefont {Henderson},\ and\ \citenamefont
  {Scuseria}}]{Wen2013ACS}%
  \BibitemOpen
  \bibfield  {author} {\bibinfo {author} {\bibfnamefont {X.-D.}\ \bibnamefont
  {Wen}}, \bibinfo {author} {\bibfnamefont {R.~L.}\ \bibnamefont {Martin}},
  \bibinfo {author} {\bibfnamefont {T.~M.}\ \bibnamefont {Henderson}},\ and\
  \bibinfo {author} {\bibfnamefont {G.~E.}\ \bibnamefont {Scuseria}},\
  }\bibfield  {title} {\bibinfo {title} {Density functional theory studies of
  the electronic structure of solid state actinide oxides},\ }\href
  {https://doi.org/10.1021/cr300374y} {\bibfield  {journal} {\bibinfo
  {journal} {Chemical Reviews}\ }\textbf {\bibinfo {volume} {113}},\ \bibinfo
  {pages} {1063} (\bibinfo {year} {2013})}\BibitemShut {NoStop}%
\bibitem [{\citenamefont {Nguyen}\ \emph {et~al.}(2018)\citenamefont {Nguyen},
  \citenamefont {Halloran}, \citenamefont {Xie}, \citenamefont {Kong},
  \citenamefont {Broholm},\ and\ \citenamefont {Cava}}]{Nguyen2018PRM}%
  \BibitemOpen
  \bibfield  {author} {\bibinfo {author} {\bibfnamefont {L.~T.}\ \bibnamefont
  {Nguyen}}, \bibinfo {author} {\bibfnamefont {T.}~\bibnamefont {Halloran}},
  \bibinfo {author} {\bibfnamefont {W.}~\bibnamefont {Xie}}, \bibinfo {author}
  {\bibfnamefont {T.}~\bibnamefont {Kong}}, \bibinfo {author} {\bibfnamefont
  {C.~L.}\ \bibnamefont {Broholm}},\ and\ \bibinfo {author} {\bibfnamefont
  {R.~J.}\ \bibnamefont {Cava}},\ }\bibfield  {title} {\bibinfo {title}
  {{Geometrically frustrated trimer-based Mott insulator}},\ }\href
  {https://doi.org/10.1103/PhysRevMaterials.2.054414} {\bibfield  {journal}
  {\bibinfo  {journal} {Phys. Rev. Mater.}\ }\textbf {\bibinfo {volume} {2}},\
  \bibinfo {pages} {054414} (\bibinfo {year} {2018})}\BibitemShut {NoStop}%
\bibitem [{\citenamefont {Sheldrick}(2015)}]{2015_Sheldrick_Acta}%
  \BibitemOpen
  \bibfield  {author} {\bibinfo {author} {\bibfnamefont {G.~M.}\ \bibnamefont
  {Sheldrick}},\ }\bibfield  {title} {\bibinfo {title} {{Crystal structure
  refinement with SHELXL}},\ }\href {https://doi.org/10.1107/S2053229614024218}
  {\bibfield  {journal} {\bibinfo  {journal} {Acta Crystallographica Section C:
  Structural Chemistry}\ }\textbf {\bibinfo {volume} {71}},\ \bibinfo {pages}
  {3} (\bibinfo {year} {2015})}\BibitemShut {NoStop}%
\bibitem [{\citenamefont {Savitzky}\ \emph {et~al.}(2018)\citenamefont
  {Savitzky}, \citenamefont {El~Baggari}, \citenamefont {Clement},
  \citenamefont {Waite}, \citenamefont {Goodge}, \citenamefont {Baek},
  \citenamefont {Sheckelton}, \citenamefont {Pasco}, \citenamefont {Nair},
  \citenamefont {Schreiber}, \citenamefont {Hoffman}, \citenamefont {Admasu},
  \citenamefont {Kim}, \citenamefont {Cheong}, \citenamefont {Anand},
  \citenamefont {Schlom}, \citenamefont {McQueen}, \citenamefont {Hovden},\
  and\ \citenamefont {Kourkoutis}}]{savitzky2018image}%
  \BibitemOpen
  \bibfield  {author} {\bibinfo {author} {\bibfnamefont {B.~H.}\ \bibnamefont
  {Savitzky}}, \bibinfo {author} {\bibfnamefont {I.}~\bibnamefont
  {El~Baggari}}, \bibinfo {author} {\bibfnamefont {C.~B.}\ \bibnamefont
  {Clement}}, \bibinfo {author} {\bibfnamefont {E.}~\bibnamefont {Waite}},
  \bibinfo {author} {\bibfnamefont {B.~H.}\ \bibnamefont {Goodge}}, \bibinfo
  {author} {\bibfnamefont {D.~J.}\ \bibnamefont {Baek}}, \bibinfo {author}
  {\bibfnamefont {J.~P.}\ \bibnamefont {Sheckelton}}, \bibinfo {author}
  {\bibfnamefont {C.}~\bibnamefont {Pasco}}, \bibinfo {author} {\bibfnamefont
  {H.}~\bibnamefont {Nair}}, \bibinfo {author} {\bibfnamefont {N.~J.}\
  \bibnamefont {Schreiber}}, \bibinfo {author} {\bibfnamefont {J.}~\bibnamefont
  {Hoffman}}, \bibinfo {author} {\bibfnamefont {A.}~\bibnamefont {Admasu}},
  \bibinfo {author} {\bibfnamefont {J.}~\bibnamefont {Kim}}, \bibinfo {author}
  {\bibfnamefont {S.-W.}\ \bibnamefont {Cheong}}, \bibinfo {author}
  {\bibfnamefont {B.}~\bibnamefont {Anand}}, \bibinfo {author} {\bibfnamefont
  {D.~G.}\ \bibnamefont {Schlom}}, \bibinfo {author} {\bibfnamefont {T.~M.}\
  \bibnamefont {McQueen}}, \bibinfo {author} {\bibfnamefont {R.}~\bibnamefont
  {Hovden}},\ and\ \bibinfo {author} {\bibfnamefont {L.~F.}\ \bibnamefont
  {Kourkoutis}},\ }\bibfield  {title} {\bibinfo {title} {{Image registration of
  low signal-to-noise cryo-STEM data}},\ }\href
  {https://doi.org/10.1016/j.ultramic.2018.04.008} {\bibfield  {journal}
  {\bibinfo  {journal} {Ultramicroscopy}\ }\textbf {\bibinfo {volume} {191}},\
  \bibinfo {pages} {56} (\bibinfo {year} {2018})}\BibitemShut {NoStop}%
\bibitem [{\citenamefont {Philipp}\ \emph {et~al.}(2022)\citenamefont
  {Philipp}, \citenamefont {Tate}, \citenamefont {Shanks}, \citenamefont
  {Mele}, \citenamefont {Peemen}, \citenamefont {Dona}, \citenamefont
  {Hartong}, \citenamefont {van Veen}, \citenamefont {Shao}, \citenamefont
  {Chen}, \citenamefont {Thom-Levy}, \citenamefont {Muller},\ and\
  \citenamefont {Gruner}}]{philipp2022very}%
  \BibitemOpen
  \bibfield  {author} {\bibinfo {author} {\bibfnamefont {H.~T.}\ \bibnamefont
  {Philipp}}, \bibinfo {author} {\bibfnamefont {M.~W.}\ \bibnamefont {Tate}},
  \bibinfo {author} {\bibfnamefont {K.~S.}\ \bibnamefont {Shanks}}, \bibinfo
  {author} {\bibfnamefont {L.}~\bibnamefont {Mele}}, \bibinfo {author}
  {\bibfnamefont {M.}~\bibnamefont {Peemen}}, \bibinfo {author} {\bibfnamefont
  {P.}~\bibnamefont {Dona}}, \bibinfo {author} {\bibfnamefont {R.}~\bibnamefont
  {Hartong}}, \bibinfo {author} {\bibfnamefont {G.}~\bibnamefont {van Veen}},
  \bibinfo {author} {\bibfnamefont {Y.-T.}\ \bibnamefont {Shao}}, \bibinfo
  {author} {\bibfnamefont {Z.}~\bibnamefont {Chen}}, \bibinfo {author}
  {\bibfnamefont {J.}~\bibnamefont {Thom-Levy}}, \bibinfo {author}
  {\bibfnamefont {D.~A.}\ \bibnamefont {Muller}},\ and\ \bibinfo {author}
  {\bibfnamefont {S.~M.}\ \bibnamefont {Gruner}},\ }\bibfield  {title}
  {\bibinfo {title} {Very-high dynamic range, 10,000 frames/second pixel array
  detector for electron microscopy},\ }\href
  {https://doi.org/10.1017/S1431927622000174} {\bibfield  {journal} {\bibinfo
  {journal} {Microscopy and Microanalysis}\ }\textbf {\bibinfo {volume} {28}},\
  \bibinfo {pages} {425} (\bibinfo {year} {2022})}\BibitemShut {NoStop}%
\bibitem [{\citenamefont {Kresse}\ and\ \citenamefont
  {Joubert}(1999)}]{Kresse1999}%
  \BibitemOpen
  \bibfield  {author} {\bibinfo {author} {\bibfnamefont {G.}~\bibnamefont
  {Kresse}}\ and\ \bibinfo {author} {\bibfnamefont {D.}~\bibnamefont
  {Joubert}},\ }\bibfield  {title} {\bibinfo {title} {From ultrasoft
  pseudopotentials to the projector augmented-wave method},\ }\href
  {https://doi.org/10.1103/PhysRevB.59.1758} {\bibfield  {journal} {\bibinfo
  {journal} {Phys. Rev. B}\ }\textbf {\bibinfo {volume} {59}},\ \bibinfo
  {pages} {1758} (\bibinfo {year} {1999})}\BibitemShut {NoStop}%
\bibitem [{\citenamefont {Kresse}\ and\ \citenamefont
  {Furthm\"{u}ller}(1996{\natexlab{a}})}]{Kresse1996a}%
  \BibitemOpen
  \bibfield  {author} {\bibinfo {author} {\bibfnamefont {G.}~\bibnamefont
  {Kresse}}\ and\ \bibinfo {author} {\bibfnamefont {J.}~\bibnamefont
  {Furthm\"{u}ller}},\ }\bibfield  {title} {\bibinfo {title} {Efficiency of
  ab-initio total energy calculations for metals and semiconductors using a
  plane-wave basis set},\ }\href {https://doi.org/10.1016/0927-0256(96)00008-0}
  {\bibfield  {journal} {\bibinfo  {journal} {Computational Materials Science}\
  }\textbf {\bibinfo {volume} {6}},\ \bibinfo {pages} {15} (\bibinfo {year}
  {1996}{\natexlab{a}})}\BibitemShut {NoStop}%
\bibitem [{\citenamefont {Kresse}\ and\ \citenamefont
  {Furthm\"{u}ller}(1996{\natexlab{b}})}]{Kresse1996b}%
  \BibitemOpen
  \bibfield  {author} {\bibinfo {author} {\bibfnamefont {G.}~\bibnamefont
  {Kresse}}\ and\ \bibinfo {author} {\bibfnamefont {J.}~\bibnamefont
  {Furthm\"{u}ller}},\ }\bibfield  {title} {\bibinfo {title} {Efficient
  iterative schemes for ab initio total-energy calculations using a plane-wave
  basis set},\ }\href {https://doi.org/10.1103/PhysRevB.54.11169} {\bibfield
  {journal} {\bibinfo  {journal} {Phys. Rev. B}\ }\textbf {\bibinfo {volume}
  {54}},\ \bibinfo {pages} {11169} (\bibinfo {year}
  {1996}{\natexlab{b}})}\BibitemShut {NoStop}%
\bibitem [{\citenamefont {Perdew}\ \emph {et~al.}(1996)\citenamefont {Perdew},
  \citenamefont {Burke},\ and\ \citenamefont {Ernzerhof}}]{Perdew1996}%
  \BibitemOpen
  \bibfield  {author} {\bibinfo {author} {\bibfnamefont {J.~P.}\ \bibnamefont
  {Perdew}}, \bibinfo {author} {\bibfnamefont {K.}~\bibnamefont {Burke}},\ and\
  \bibinfo {author} {\bibfnamefont {M.}~\bibnamefont {Ernzerhof}},\ }\bibfield
  {title} {\bibinfo {title} {Generalized gradient approximation made simple},\
  }\href {https://doi.org/10.1103/PhysRevLett.77.3865} {\bibfield  {journal}
  {\bibinfo  {journal} {Phys. Rev. Lett.}\ }\textbf {\bibinfo {volume} {77}},\
  \bibinfo {pages} {3865} (\bibinfo {year} {1996})}\BibitemShut {NoStop}%
\bibitem [{\citenamefont {Dudarev}\ \emph {et~al.}(1998)\citenamefont
  {Dudarev}, \citenamefont {Botton}, \citenamefont {Savrasov}, \citenamefont
  {Humphreys},\ and\ \citenamefont {Sutton}}]{Dudarev1998}%
  \BibitemOpen
  \bibfield  {author} {\bibinfo {author} {\bibfnamefont {S.~L.}\ \bibnamefont
  {Dudarev}}, \bibinfo {author} {\bibfnamefont {G.~A.}\ \bibnamefont {Botton}},
  \bibinfo {author} {\bibfnamefont {S.~Y.}\ \bibnamefont {Savrasov}}, \bibinfo
  {author} {\bibfnamefont {C.~J.}\ \bibnamefont {Humphreys}},\ and\ \bibinfo
  {author} {\bibfnamefont {A.~P.}\ \bibnamefont {Sutton}},\ }\bibfield  {title}
  {\bibinfo {title} {{Electron-energy-loss spectra and the structural stability
  of nickel oxide: An LSDA+U study}},\ }\href
  {https://doi.org/10.1103/PhysRevB.57.1505} {\bibfield  {journal} {\bibinfo
  {journal} {Phys. Rev. B}\ }\textbf {\bibinfo {volume} {57}},\ \bibinfo
  {pages} {1505} (\bibinfo {year} {1998})}\BibitemShut {NoStop}%
\bibitem [{\citenamefont {Ganose}\ \emph {et~al.}(2021)\citenamefont {Ganose},
  \citenamefont {Searle}, \citenamefont {Jain},\ and\ \citenamefont
  {Griffin}}]{Ganose2021}%
  \BibitemOpen
  \bibfield  {author} {\bibinfo {author} {\bibfnamefont {A.~M.}\ \bibnamefont
  {Ganose}}, \bibinfo {author} {\bibfnamefont {A.}~\bibnamefont {Searle}},
  \bibinfo {author} {\bibfnamefont {A.}~\bibnamefont {Jain}},\ and\ \bibinfo
  {author} {\bibfnamefont {S.~M.}\ \bibnamefont {Griffin}},\ }\bibfield
  {title} {\bibinfo {title} {{IFermi: A python library for Fermi surface
  generation and analysis}},\ }\href {https://doi.org/10.21105/joss.03089}
  {\bibfield  {journal} {\bibinfo  {journal} {Journal of Open Source Software}\
  }\textbf {\bibinfo {volume} {6}},\ \bibinfo {pages} {3089} (\bibinfo {year}
  {2021})}\BibitemShut {NoStop}%
\end{thebibliography}%

\end{document}